\documentclass[aps,prx,twocolumn,superscriptaddress,showpacs]{revtex4-2}
\usepackage[dvipdfmx]{graphicx}
\usepackage{amsmath,amsbsy,amssymb}
\usepackage{bm}
\usepackage{mathrsfs}
\usepackage{ulem}
\usepackage{textgreek}
\usepackage{multirow}
\usepackage{color}

\newcommand{\diff}{\mathrm{d}}

\newcommand{\trace}{\mathrm{Tr}\,}
\newcommand{\imu}{\mathrm{i}}

\newcommand{\ua}{\uparrow}
\newcommand{\da}{\downarrow}
\newcommand{\dg}{\dagger}
\newcommand{\la}{\langle}
\newcommand{\ra}{\rangle}
\newcommand{\al}{\alpha}
\newcommand{\sg}{\sigma}
\newcommand{\gm}{\gamma}

\begin{document}

\title{
Multipole representation for anisotropic Coulomb interactions 
}

\author{Shoma Iimura$^*$}
\affiliation{Department of Physics, Saitama University, Sakura-ku, Saitama 338-8570, Japan}

\author{Motoaki Hirayama$^*$}
\affiliation{Quantum-Phase Electronics Center, The University of Tokyo, 
Bunkyo-ku,
Tokyo 113-8656, Japan}
\affiliation{RIKEN Center for Emergent Matter Science (CEMS), Wako, Saitama 351-0198, Japan}

\author{Shintaro Hoshino$^\dg$}
\affiliation{Department of Physics, Saitama University, Sakura-ku, Saitama 338-8570, Japan}

\date{\today}

\begin{abstract}
Multipole representation is proposed for the anisotropic Coulomb interactions in solids.
Any local interactions can be expressed as the product of two multipole operators, and the 
interaction parameters are systematically classified based on the point group symmetry.
The form of the multipole interactions are restricted not only by the symmetry and Hermiticity but also by the spatial structure of the interaction, which is closely related to the presence or absence of the odd-rank multipoles.
As an exemplary demonstration, the screened Coulomb interaction for SrVO$_3$ is considered, where only a few parameters are necessary for its description.
By comparing it with the unscreened version, the totally symmetric $A_1$ representation is found to be strongly suppressed, but the $A_1$ component still gives a dominant contribution for the anisotropic part of the interaction.
The anisotropic interactions are also applied to the localized two $f$-electron wave functions, which give the same-order contribution as the one-body level splitting estimated by the band structure calculation.
\end{abstract}

\maketitle

The interactions among electrons cause a variety of intriguing phenomena in strongly correlated systems.
In solids, the tight-binding model is frequently used for the description of the correlated electronic states, where the Coulomb interactions can be expressed in the second quantized Hamiltonian by the product of the four creation/annihilation operators of electrons.
Although for most cases the Hubbard model with the on-site interaction is enough for the phenomena of interest, the multiorbital nature is still necessary to be considered for almost all the strongly correlated materials including $d$-electron systems such as iron-based superconductors, heavy-electron materials, molecular-based conductors \cite{review}.

Usually, the Coulomb interaction is considered as the one in the spherical limit, where only a few parameters are necessary \cite{Condon_book}.
In the presence of the multiorbital effects in solids, however, a discontinuous point group symmetry at the correlated site complicates the spatial structure of the interaction.
The Coulomb interaction in solids
has been systematically studied for a spherical interaction with cubic crystalline field \cite{Coury16,Sugano_book}. 
The more general interactions have also been considered by B\"{u}nemann and Gebhard, where the interaction parameters are classified based on point group symmetries \cite{Bunemann17}.
In the present paper, we propose a simple representation in terms of multipole operator which was introduced for description of the local degrees of freedom of $f$ electrons \cite{Ohkawa83, Shiina97, Kuramoto00, Santini00, Kiss05, Takimoto05, Kusunose08, Kuramoto09, Haule09, Ikeda12} in terms of total angular momentum $J$.
The multipoles have also been used for analyzing the two-particle Green functions \cite{Tazai19}, and applied to the other systems beyond the scope of $f$-electrons \cite{Kugel72, Khaliullin13, Suzuki17, Hayami18}.
As demonstrated in this paper, the multipole representation makes it simpler to consider the anisotropic Coulomb interactions in solids with discrete point group symmetries.
There are even-rank and odd-rank multipoles as classified by the time-reversal symmetry (TRS), and we show that the spatial structure of the interaction is closely connected to the presence or absence of the odd-rank multipoles in the interaction.
The proposed scheme is used for analyzing the complicated effective interactions in the first-principle calculations \cite{Imada10,Miyake10,Misawa11,Misawa12,Misawa14,Hirayama15,Hirayama18,Hirayama19,Ohgoe20}.

The local Coulomb interaction in solids is written in a general form as
\begin{align}
\mathscr H_{\rm C} &= \frac 1 2
\int \diff \bm r_1 \diff \bm r_2 \diff \bm r_3 \diff \bm r_4
\nonumber \\
&\ \ \ 
\times U(\bm r_1,\bm r_2, \bm r_3, \bm r_4)
: n(\bm r_1,\bm r_3) n(\bm r_2,\bm r_4) :
,
\end{align}
where $n(\bm r,\bm r') = \sum_{\sg} \psi^\dg_{\sg}(\bm r) \psi_\sg(\bm r')$ with the annihilation operator $\psi_\sg (\bm r)$ of electrons with spin $\sg$ ($=\ua,\da$) .
The colon ($:$) symbol makes the expression normal ordering, i.e., the creation operators are placed left and annihilation operators right with the consideration of anticommutation relation \cite{Coury16}.
For specific principal and azimuthal quantum numbers ($n,\ell$) of localized electrons at an atom, it is expressed as
\begin{align}
\mathscr H_{\rm C} &= \frac 1 2 \sum_{m_1m_2m_3m_4} U_{m_1m_2m_3m_4} : n_{m_1m_3} n_{m_2m_4} :
, \label{eq:coulomb}
\end{align}
where the density matrix operator is defined by $n_{mm'} = \sum_\sg c^\dg_{m\sg} c_{m'\sg}$ with the electron annihilation operator $c_{m\sg}$ of a magnetic quantum number $m$.

For a spherically symmetric case, it is well known that the above matrix element can be expressed by the Slater-Condon parameters, which are denoted as $F^{k}$ where $k$ is a rank~\cite{Condon_book}.
Owing to the symmetry of the Gaunt coefficient, the odd $k$ terms can be set as zero and only $\ell+1$ parameters are needed ($k=0,2,\cdots,2\ell$).
In solids, on the other hand, the continuous symmetry does not exist, and therefore the matrix elements have much complicated structure as seen in, e.g., Ref.~\cite{Hirayama19}.

The central idea of this paper is to rewrite the Coulomb interaction in terms of multipole operators, which are defined by
$M_{\xi} = \sum_{mm'\sg} c_{m\sg}^\dg O^\xi_{mm'} c_{m'\sg} $
where $\xi$ is the index for the multipole and $\hat O^\xi$ is a Hermitian $(2\ell+1)\times (2\ell+1)$ matrix.
These are constructed through the combination of angular momentum operators \cite{Kusunose08}.
The number of the full set of matrices is also $(2\ell+1)^2$.
Hence the series of the matrices is regarded as complete, and any matrix can be expanded by these matrices.
More specifically, the matrices satisfy the relations
$\trace \hat O^{\xi} \hat O^{\xi'} = (2\ell + 1) \delta_{\xi\xi'}$ and 
$\sum_\xi O_{m_1m_2}^{\xi} O_{m_3m_4}^{\xi} = (2\ell + 1) \delta_{m_1m_4}\delta_{m_2m_3}$.
These relations can be understood by considering the most simple situation with $2\times 2$ matrices, where the three Pauli matrices and identity matrix are involved.
We can transform the density matrix operator $n_{mm'}$ into the multipole operators, and the interaction is written as
\begin{align}
\mathscr H_{\rm C} &= \frac 1 2 \sum_{\xi\xi'} I(\xi,\xi') : M_{\xi} M_{\xi'} :
, \label{eq:coulomb_multipole}
\end{align}
where the relation 
$I(\xi,\xi') = I(\xi',\xi) = I^*(\xi,\xi')$ holds.
Starting from Eq.~\eqref{eq:coulomb}, we can perform the multipole expansion uniquely with a given set of matrices.
This is the most general expression for the local Coulomb interaction in the absence of the spin-orbit coupling in the interaction terms \cite{comment1}.

In solids, the multipole matrices are classified by the rank, which corresponds to the angular momentum of multipoles, and also by the irreducible representation under a given point group \cite{Kusunose08}.
The complete set of the multipoles are explicitly given in the literatures \cite{Kusunose08,Hayami18,Kusunose20}, and here we utilize them for 
description of interactions  (See Supplementary Material (SM) A \cite{suppl} for more details).
The types of the multipoles 
are obtained by considering the irreducible decomposition for each rank \cite{Sugano_book, Dresselhaus_book}, and we summarize the results in Tab.~\ref{tab:decomp} for the cubic point group. 
The index for multipoles is then written as $\xi = (k,\Gamma,\al)$,
where
$\Gamma$ identifies the type
listed in Tab.~\ref{tab:decomp}
and $\al$ distinguishes the degenerate components belonging to $\Gamma$. 
Note that $\Gamma$ is implicitly dependent on $k$ (see Tab.~\ref{tab:decomp}), and also $\al$ dependent on $\Gamma$.
The multipole interaction now becomes
\begin{align}
\label{eq:interaction}
\mathscr H_{\rm C} &= \frac 1 2 \sum_{k\Gamma}\sum_{k'\Gamma'} I(k,k';\Gamma,\Gamma')
\sum_{\al} : M_{k\Gamma\al} M_{k'\Gamma'\al} :
.
\end{align}
Because of selection rules in the group theory, the multipoles with different irreducible representations do not interact \cite{suppl}.
However, this does not mean the interaction is diagonal with respect to $\Gamma$: for example, the interaction parameter $I(2,6; T_2 , T_2^{\rm a})$ can be finite.
This property is also checked by using the concrete expressions for multipoles given in Refs.~\cite{Kusunose08,Hayami18}.
While we restrict ourselves to a fixed $\ell$ case, the extension for the parity mixing between the different orbital angular momenta is also possible with using suitable multipole basis \cite{Kusunose20}.
The above multipole representation can be identified with the familiar Slater-Condon parameters $F^{k}$ in the spherical limit.
In this case $I$ is dependent only on the rank $k$ of the multipoles, and 
there is a simple correspondence $I(k,k') \propto F^{k} \delta_{kk'}$ \cite{suppl}.

For a time-reversal symmetric system,
the even- and odd-rank multipoles do not mix, since the odd-rank multipoles are odd under the time-reversal transformation.
Actually, the appearance of even-rank or odd-rank multipoles are also closely related to the functional form of $U(\bm r_1,\bm r_2,\bm r_3, \bm r_4)$.
Usually, the interaction $U(\bm r_1,\bm r_2,\bm r_1,\bm r_2)$ at two spatial point is considered.
In this case, the interaction includes only the even-rank multipoles (See SM B \cite{suppl}), and hence it does not have an ability to describe the TRS breaking in interaction terms.
The complexity enters when we consider the more general case $U(\bm r_1,\bm r_2,\bm r_1,\bm r_4)$, which can be realized by considering the static three-point vertex correction to the Coulomb interaction.
However, even in this case, we do not have odd-rank multipoles for the time-reversal symmetric system \cite{suppl}.
Thus, only the case of $U(\bm r_1,\bm r_2,\bm r_3, \bm r_4)$ with static four-point vertex corrections induces the odd-rank multipole interactions.
We note that, for a TRS broken system, the three-point function can have the odd-rank multipoles coupled to even-rank multipoles, but the two-point function does not have an ability to describe the TRS breaking.
In this way, the appearance or disappearance of odd-rank multipoles is connected with the spatial structure of the interaction.

Below, we consider the three specific cases as a demonstration of multipole representation of the Coulomb interactions.

\begin{table}[t]
\centering
\begin{tabular}{ccl}
\hline
rank of multipoles & \ \  & Type of multipoles ($\Gamma$) \\
\hline
$k=0$ & & $A_1$ \\
$k=1$ & & $T_1$ \\
$k=2$ & & $E+T_2$ \\
$k=3$ & & $A_2+T_1+T_2$ \\
$k=4$ & & $A_1+E+T_1+T_2$ \\
$k=5$ & & $E+T_1^{\rm a}+T_1^{\rm b}+T_2$ \\
$k=6$ & & $A_1+A_2+E+T_1+T_2^{\rm a}+T_2^{\rm b}$ \\
\hline
\end{tabular}
\caption{
List of multipoles for each rank classified by the irreducible representations under the cubic $O_h$ point group symmetry.
See Ref.~\cite{Kusunose08} for the concrete form of the multipoles.
}
\label{tab:decomp}
\end{table}

\begin{figure*}[t]
\begin{center}
\includegraphics[width=180mm]{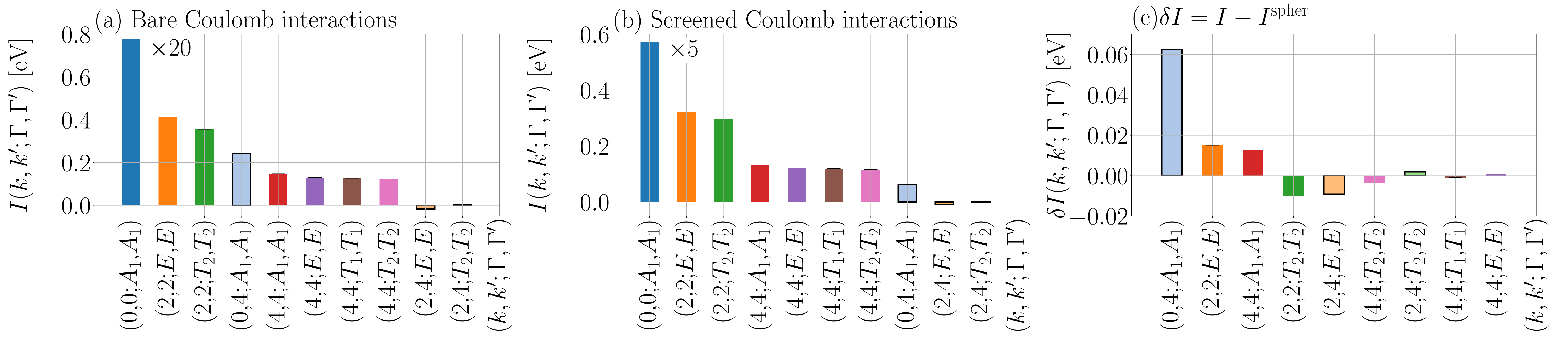}
\caption{
Local Coulomb interaction parameters for SrVO$_3$.
Multipole representation for (a) unscreened interaction and (b) screened interaction calculated by cRPA are shown.
The value of $I(0,0;A_1,A_1)$ needs to be multiplied by 20 in (a) and 5 in (b).
In (c), the deviation from the spherical limit $\delta I = I - I^{\rm spher}$ for the screened Coulomb interaction in (b) is plotted.
}
\label{fig:svo}
\end{center}
\end{figure*}

{\it Application to $p$ electrons.---}
In order to have intuition for multipole interactions, let us first consider the $p$-electrons ($\ell=1$) with real wave function basis.
The results are also applicable to the $t_{2g}$ orbitals of $d$-electrons \cite{Georges13} and $t_{1u}$ orbitals of fulleride materials \cite{Nomura16}.
The interaction is usually parametrized as the Slater-Kanamori interaction \cite{Slater_book, Kanamori63,Castellani78}: $U_{\gm\gm\gm\gm}=U$, $U_{\gm\gm'\gm\gm'}=U'$, $U_{\gm\gm'\gm'\gm} = U_{\gm\gm\gm'\gm'} =J$ ($\gm\neq \gm'$) where $\gm$ represents $p_{x,y,z}$ orbitals instead of the magnetic angular momentum $m$ ($=0,\pm 1$).
The relations to the multipole representation are identified as 
$I(0,0;A_1,A_1) = \frac{1}{3} (U+2 U')$, $I(2,2;E,E)= \frac{1}{3} (U-U')$, $I(2,2;T_2,T_2) = \frac{2}{3} J$ and $I(1,1;T_1 , T_1) = 0$.
The cubic symmetry is reflected in the difference between $I(2,2;E,E)$ and $I(2,2;T_2,T_2)$, and in the spherically symmetric case we have the relation $I(2,2;E,E) = I(2,2;T_2,T_2)$ identical to the well-known condition $U'=U-2J$.
We emphasize that only in this spherical case can the Slater-Condon parametrization ($F^{0,2}$) be used.
The odd-rank multipole with $\Gamma=T_1$ is absent in this case which is related to the fact that the Slater-Kanamori parameterization is based on the interactions at two spatial points.

{\it Application to $d$ electrons.---}
Next we consider the actual materials.
For an exemplary demonstration, we take the concrete material SrVO$_3$.
We calculate the electronic band structure of SrVO$_3$ from the local density approximation (LDA) of the density functional theory (DFT) \cite{methfessel,ceperley} and construct the maximally localized Wannier function of the V $3d$ orbitals hybridized with the O $2p$ orbitals.
We also perform the constrained random phase approximation (cRPA) technique to calculate the screened interaction for the Wannier functions, which reflects the screening effects in the cubic crystal \cite{aryasetiawan04,hirayama13}.
The detail of the \textit{ab initio} calculation is shown in SM C \cite{suppl}.

Figure~\ref{fig:svo} shows the multipole interactions for $3d$-electrons at the V site with $O_h$ point group symmetry, where the values are listed in a descending order with respect to the absolute values.
Since the numerical errors are in general included in the raw data \cite{suppl}, we symmetrize the interactions, but the errorbars are invisible for our data.
The bare Coulomb interactions are shown in (a).
We can see, for instance, the difference between $I(k,k;E,E)$ and $I(k,k;T_2,T_2)$ which reflects the cubic symmetry.
It is notable that only few parameters are relevant to describe the Coulomb interaction tensor.
The screened interactions are also shown in Fig.~\ref{fig:svo}(b), where the $A_1$ components are much suppressed, while the others remain almost unchanged.
This is understood as follows: 
The electronic charge is responsible for the screening to reduce the Coulomb interaction, and the charge component is represented as totally symmetric representation $A_1$.
Because of the cubic symmetry, the different ranks for $A_1$ are mixed and are much influenced by screening.
Intuitively, the parameter $I(0,4;A_1,A_1)$ may be interpreted as that the spherical deformation of charge is accompanied by the cubic deformation in the solids with $O_h$ symmetry.

We note that the odd-rank multipoles are absent in the cRPA results.
This is consistent with the fact that the RPA calculation is performed for a fixed single wave vector $\bm q$ of the effective interaction 
where no three-point vertex correction is considered.
The effective interaction then originates from a function of the two spatial points and the odd-rank multipoles do not appear as discussed before.
Since the spatial inversion does not change the local interaction term for a fixed $\ell$ \cite{suppl}, the absence of the odd-rank multipoles indicates that the local cRPA interaction is always time-reversal and inversion symmetric.

The result of the screened Coulomb interactions shown in Fig.~\ref{fig:svo}(b) can be represented as the sum of the spherically symmetric part $I^{\rm spher}$ plus its deviation, which is suitable for examining the contribution from the discrete cubic symmetry.
The spherical part of the interaction is extracted as
$I^{\rm spher}(k) = \frac{1}{2k+1} \sum_{\Gamma}d_{\rm \Gamma }I(k,k;\Gamma,\Gamma)$
where $d_{\Gamma} = \sum_{\al}1$ is the number of degeneracy.
This is used for the definition of the cubic deviation $\delta I = I - I^{\rm spher}$ shown in Fig.~\ref{fig:svo}(c).
Here the dominant component is $I(0,4;A_1,A_1)$ as compared to the others, even though it has been much reduced by the screening effect.
The second largest one is the rank-$2$ with non-$A_1$ representation, and the other interaction values are basically decrease as the rank increases.
The present results thus indicate that the dominant contributions for the anisotropic part are given by the $A_1$ component plus rank-$2$ component.

Whereas the data in Fig.~\ref{fig:svo} are sufficiently accurate, the more complex materials may produce the larger numerical errors.
The quality of data can be improved by using the symmetries, and if the interaction originates from spatial two-point functions as in cRPA, the odd-rank multipoles can also be dropped, which is not a requirement of symmetry.

\begin{figure*}[t]
\begin{center}
\includegraphics[width=160mm]{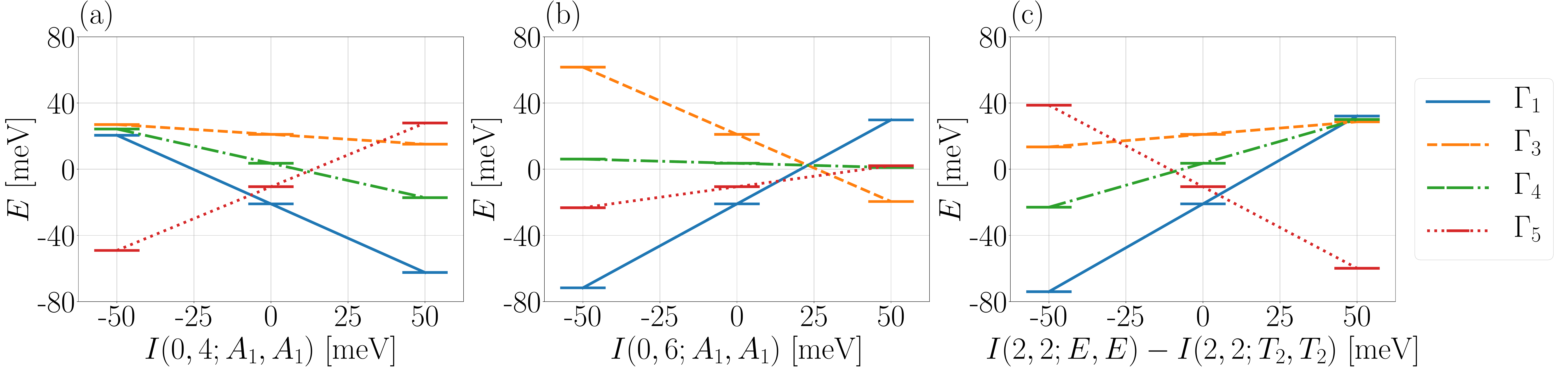}
\caption{
Anisotropic interaction effects on the crystal field energy levels of $f^2$ wave function.
The types of interaction parameters are (a) $I(0,4;A_1,A_1)$, (b) $I(0,6;A_1,A_1)$, and (c) $I(2,2;E,E)-I(2,2;T_2,T_2)$.
The level scheme of the local one-body potential calculated for UBe$_{13}$ is shown at $I=0$ and the deviation enters for $I\neq 0$.
}
\label{fig:ube13}
\end{center}
\end{figure*}

{\it Application to $f$ electrons.---}
We further apply the anisotropic multipole interactions to the localized $f$-electron wave function under the cubic crystalline field \cite{Lea62}.
We consider the two $f$-electron ($f^2$) wave functions realized in Pr- and U-based materials, for which the interaction effects are relevant.
By considering the spherical part of the Coulomb interaction and the spin-orbital coupling (Hund's rule), we obtain the ground state $J=4$ multiplet $|M\ra$ ($M \in [-J,J]$) (SM D \cite{suppl}).
We now demonstrate that the anisotropic interaction affects the wave function and modifies the energy level structure.
We take UBe$_{13}$ as an example, which
shows the robust non-Fermi liquid behavior and unconventional superconductivity, and the multichannel Kondo effects have been suspected as possible origins \cite{Cox87,Cox98}.
In the multichannel Kondo effect scenario, the realization of non-Kramers $\Gamma_3$ doublet ground state is a necessary condition for a robust non-Fermi liquid.
In our previous works, we focus on the fact that the conduction electrons of UBe$_{13}$ can be seen as compensated metal and propose possible scenario for the unconventional superconductivity \cite{Iimura19,Iimura20}.
On the other hand, there is another possibility that the competition between Kondo singlet and crystalline field $\Gamma_1$ singlet leads also to the non-Fermi liquids \cite{Yotsuhashi05,Nishiyama10}.
With these backgrounds, we consider the effect of the anisotropic multipole interaction on the $f^2$ wave functions and see what kind of the ground state is favored by multipole interactions.

For the estimation of the local one-body level splitting , we calculate the band structures of UBe$_{13}$ and 
find the local on-site potentials for $f$ electrons \cite{suppl}.
Since Fig.~\ref{fig:svo} shows that the dominant contribution is the $A_1$ type involving the rank 0 component, we consider the multipole interactions $I(0,4;A_1,A_1)$, $I(0,6;A_1,A_1)$, and also the second largest one $I(2,2;E,E) - I(2,2;T_2,T_2)$ in Fig.~\ref{fig:svo}(c).
The magnitude of the typical values are estimated as $I(0,4;A_1,A_1)/I(0,0;A_1,A_1) \simeq 0.02$ from Fig.~\ref{fig:svo}(b,c).
Assuming that $I(0,0;A_1,A_1)$ ($=F^0$) is nearly 2.5eV for U atom \cite{Marel88}, we consider the range $|I| < 50$meV.
The results are shown in Fig.~\ref{fig:ube13}, where the anisotropic interactions are included by the first-order perturbation theory \cite{suppl}.
The crystalline field $\Gamma_1$ singlet is the ground state without multipole interactions, and
once the anisotropy is introduced, we find $\Gamma_1$, $\Gamma_3$ and $\Gamma_5$ ground states depending on the parameters.
Thus the cubic anisotropy of the interaction can substantially modify the crystal field structure determined by the non-correlated parts.

This scheme is applicable to any types of materials, and hence our results show that we need to be careful about the anisotropic interaction effects when one determines  the multiple $f$-electron wave functions from a microscopic point of view.
For the one-body part of the Hamiltonian, the energy spectrum is expected to be accurately described by the first-principle band structure calculations, but the interaction effect considered in this paper is a correlation effect which is not included in the band-structure calculations.
We note that this is not true for the $f^1$ in Ce and its hole analog, $f^{13}$ in Yb, since the interaction effects for the localized crystal field levels are irrelevant.

To summarize, we have proposed a systematic and simple way to express the Coulomb interaction in solids by using the multipole operators.
The interaction parameters are restricted by the symmetries, and we have also identified that the spatial structure of the interaction functional form is closely connected to the presence or absence of odd-rank multipoles.
The multipole representation can be utilized for examining the structure of the complicated cRPA interactions and for studying the crystal field ground states of the localized correlated electrons.
Whereas we focus on the cubic crystal in the present paper, in principle, the formulation with multipoles can be applied to any local interactions including molecules and quasicrystals, and can also be generalized for the inter-site interactions.

\section*{Acknowledgement}
We are grateful to Tatsuya Miki for useful discussions.
This work was supported by JSPS KAKENHI Grants 
No.~JP18K13490,
No.~JP18H01176 and
No.~JP19H01842.

\clearpage
\appendix

\makeatletter
\renewcommand{\theequation}{S\arabic{equation}}
\renewcommand{\thefigure}{S\arabic{figure}}
\renewcommand{\thetable}{S\arabic{table}}
\renewcommand{\thepage}{S\arabic{page}}
\makeatother

\setcounter{page}{1}
\setcounter{equation}{0}
\setcounter{table}{0}
\setcounter{figure}{0}

\noindent
{\bf SUPPLEMENTARY MATERIAL FOR 
\\
``Multipole representation for anisotropic Coulomb interactions ''
}
\\[2mm]
S. Iimura, M. Hirayama, and S. Hoshino
\\[2mm]
(Dated: \today)

\section*{SM A. Complete orthonormal basis for matrix representations}
Following the procedure given in Ref.~\cite{Kusunose08}, we construct the complete orthonormal basis for the matrices.
Since the full list is shown in the literatures \cite{Kusunose08, Hayami18, Kusunose20}, we do not list the full set of matrices but we only show a few of them.
First of all, we define the angular momentum $(2\ell+1)\times (2\ell+1)$ matrices $\hat L_{x,y,z}$, and their combination leads to the complete set of matrices $\hat O^{k,\Gamma,\al}$.
More concretely, we introduce the diagonal matrix $\hat L_z = {\rm diag\,}(\ell, \ell-1, \cdots, -\ell)$ by determining the quantized axis, and then the $x$ and $y$ components are constructed as they satisfy the commutation relations $[\hat L_\mu, \hat L_\nu] = \imu \sum_\lambda \epsilon_{\mu\nu\lambda} \hat L_{\lambda}$, where $\epsilon_{\mu \nu \lambda}$ is the antisymmetric tensor. 
The trivial one is the rank $0$ matrix
\begin{align}
\hat O^{0,A_1} = \hat 1
,
\end{align}
where we have omitted the index $\al$ for the one-dimensional representations.
The rank $1$ matrices are equivalent to the angular momentum matrix.
Noting that the trace of squared matrix is normalized to $2\ell+1$, we obtain
\begin{subequations}
\begin{align}
\hat O^{1,T_1,1} &= \sqrt{\frac{3}{\ell (\ell+1)}} \hat L_{x} , \\
\hat O^{1,T_1,2} &= \sqrt{\frac{3}{\ell (\ell+1)}} \hat L_{y} ,\\
\hat O^{1,T_1,3} &= \sqrt{\frac{3}{\ell (\ell+1)}} \hat L_{z} .
\end{align}
\end{subequations}
The rank $k=2$ operators are made from the combinations of $\hat L_{x,y,z}$.
We utilize the polynomials $x^2-y^2$, $3z^2-r^2$ ($r^2=x^2+y^2+z^2$), $xy$, $yz$, $zx$ for the rank 2 representation, and replace them by the angular momentum matrix to obtain the rank 2 matrices.
In order to make it Hermitian, we symmetrize the expression and obtain
\begin{subequations}
\begin{align}
\hat O^{2,E,1} &\propto 3 \hat L_z^2 - \hat {\bm L}^2 , \\
\hat O^{2,E,2} &\propto \hat L_x^2 - \hat L_y^2 , \\
\hat O^{2,T_2,1} &\propto \hat L_x\hat L_y + \hat L_y\hat L_x , \\
\hat O^{2,T_2,2} &\propto \hat L_y\hat L_z + \hat L_z\hat L_y , \\
\hat O^{2,T_2,3} &\propto \hat L_z\hat L_x + \hat L_x\hat L_z .
\end{align}
\end{subequations}
Note that we need to normalize the expressions.
Repeating the same procedure for the higher order ranks, we obtain the complete set of the $(2\ell+1)^2$ matrices.

\begin{table}[t]
\centering
\begin{tabular}{ccl}
\hline
rank & \ \  & Type of multipoles ($\Gamma$) \\
\hline
$k=0$ & & $A_1$ \\
$k=1$ & & $A_2+E_1$ \\
$k=2$ & & $A_1+E_1+E_2$ \\
$k=3$ & & $A_2+B_1+B_2+E_1+E_2$ \\
$k=4$ & & $A_1+B_1+B_2+E_1+E_2^{\rm a}+E_2^{\rm b}$ \\
$k=5$ & & $A_2+B_1+B_2+E_1^{\rm a}+E_1^{\rm b}+E_2^{\rm a}+E_2^{\rm b}$ \\
$k=6$ & & $A_1^{\rm a}+A_1^{\rm b}+A_2+B_1+B_2+E_1^{\rm a}+E_1^{\rm b}+E_2^{\rm a}+E_2^{\rm b}$ \\
\hline
\end{tabular}
\caption{
List of multipoles for each rank classified by the irreducible representations under the hexagonal $D_{6h}$ point group symmetry.
}
\label{tab:decomp2}
\end{table}

While we focus on the cubic point group, its matrix basis is also suitable for the tetragonal case.
The multipoles for hexagonal point group are also constructed with the other polynomials \cite{Hayami18}, which are listed in Tab.~\ref{tab:decomp2}.

\section*{SM B. Properties of Coulomb interactions}

\subsection{Simplification of interaction}
We begin with the general interaction form given by 
\begin{align}
\mathscr H_{\rm C} &= \frac 1 2 \sum_{k\Gamma\al}\sum_{k'\Gamma'\al'} \tilde I(k,k';\Gamma,\Gamma';\al,\al')
: M_{k\Gamma\al} M_{k'\Gamma'\al'} : .
\label{eq:diagonalization}
\end{align}
Here, we show that the interaction parameter $I$ is diagonal with respect to $\al$ and $\al'$.
First we consider a set of symmetry operations $g$, which keeps the Hamiltonian invariant.
The corresponding transformation is represented as $\mathscr D(g)$.
For a fixed $\ell$, we do not have to consider the inversion symmetry (see SM B3), and we consider only the rotation written in the form $\mathscr D(g) = \exp \big[ - \imu \bm \theta(g) \cdot \bm L
\big]$ with the angular momentum vector $\bm L = \sum_{mm'\sg}c^\dg_{m\sg} \hat {\bm L}_{mm'} c_{m'\sg}$ where the matrix $\hat {\bm L} = ({\hat L}_x,{\hat L}_y,{\hat L}_z)$ is introduced in SM A.
The fermion annihilation operator is transformed as
\begin{align}
\label{eq:rotation}
\mathscr D(g) c_{m\sg} \mathscr D(g^{-1}) &= \sum_{m'}\exp \big[
\imu \bm \theta(g) \cdot \hat {\bm L}
\big]_{mm'} c_{m'\sg}
.
\end{align}
Correspondingly the multipole operator is transformed as
\begin{align}
	\label{eq:rotation_IR}
	\mathscr D (g) M_{k\Gamma\al} \mathscr D (g^{-1}) = \sum_{\al'}M_{k\Gamma\al'}  [ \hat D_{{\rm ir}(\Gamma)} (g) ]_{\al' \al},
\end{align}
where $\hat D_{{\rm ir(\Gamma)}} (g)$ is the irreducible representation matrix for the operation $g$ with the dimension $d_{\Gamma} = \sum_\al 1$.
The symbol `${\rm ir}$' extracts the irreducible representation to which $\Gamma $ belongs:
for example, for $k=6$ in Tab.~\ref{tab:decomp} of the main text, we have ${\rm ir}(T_2^{\rm a}) = {\rm ir}(T_2^{\rm b}) = T_2$.
Since the interaction in Eq.~\eqref{eq:diagonalization} is invariant under the transformation $g$, 
we obtain the relation
\begin{align}
&\sum_{\al_1 \al_2} \tilde I(k,k';\Gamma,\Gamma';\al_1,\al_2) 
[ \hat D_{\rm ir(\Gamma )} (g) ]_{\al  \al_1} 
[ \hat D_{\rm ir(\Gamma')} (g) ]_{\al' \al_2} \nonumber \\
&=   \tilde I(k,k';\Gamma,\Gamma';\al,\al')
.
\end{align}
For a given $(k,k';\Gamma,\Gamma')$, 
this equation can be viewed as the matrix relation:
\begin{align}
&\hat D_{\rm ir(\Gamma)} (g)  \hat{A}  = \hat{A} \hat D_{\rm ir(\Gamma')} (g)
,
\end{align}
where $A_{\al\al'} = \tilde I(k,k';\Gamma,\Gamma';\al,\al')$ and 
$\hat{D}_{\rm ir(\Gamma)} \in \mathbb R$.
Using the Schur's lemma in the group theory \cite{Sugano_book,Dresselhaus_book}, 
$\hat{A}$ becomes zero matrix for ${\rm ir (\Gamma) } \neq {\rm ir (\Gamma')}$ and identity matrix for ${\rm ir (\Gamma) } = {\rm ir (\Gamma')}$.
Thus we obtain 
\begin{align}
\tilde I(k,k';\Gamma,\Gamma';\al,\al') &= I(k,k';\Gamma,\Gamma') \delta_{\rm ir(\Gamma),ir(\Gamma')} \delta_{\al\al'} 
, \label{eq:simple_I}
\end{align}
which corresponds to Eq.~\eqref{eq:interaction} of the main text. 
We note that 
 $\Gamma \neq \Gamma'$ is allowed.
Equation~\eqref{eq:simple_I} can be used for a check of the matrix basis classified by the irreducible representations.

\subsection{Symmetrization of numerical data}

The numerical data for the screened Coulomb interaction estimated by the cRPA contain the errors.
In order to remove the errors, we exploit the projection of the Hamiltonian onto the $A_1$ representation.
We consider the $O_h$ point group for SrVO$_3$, which has symmetry operations $g$.
The projection is performed for the interaction term of the Hamiltonian as
\begin{align}
\mathscr H_{\rm C}
&= \frac{1}{\sum_g 1} \sum_g \mathscr D(g) \mathscr H_{\rm C}^{\rm raw} \mathscr D(g^{-1})
,
\end{align}
where $\mathscr H_{\rm C}^{\rm raw}$ is constructed from the raw numerical data and includes numerical errors.
Thus we obtain the interaction parameter $I(k,k';\Gamma,\Gamma')$.

The information of the raw data can be used for the estimation of the numerical errors.
The original data for the Coulomb interaction includes the errors and its multipole representation is 
\begin{align}
\mathscr H^{\rm raw}_{\rm C} &= \frac 1 2 \sum_{k\Gamma\al} \sum_{k'\Gamma'\al'} 
I^{\rm raw}(k,k';\Gamma,\Gamma';\al,\al') 
\nonumber \\
&\hspace{20mm} \times \, :M_{k\Gamma\al} M_{k'\Gamma'\al'}:
\end{align}
For our case, the interaction parameters are symmetrized as $I^{\rm raw}(k,k';\Gamma,\Gamma',\al,\al') \to I(k,k';\Gamma,\Gamma') \delta_{\al\al'}$, for which 
we can estimate the error $\varDelta I$ by 
\begin{align}
\varDelta I(k,k';\Gamma,\Gamma')^2 &=  
\frac{1}{d_\Gamma d_{\Gamma'}} \sum_{\al \al'} \big[ I^{\rm raw} (k,k';\Gamma,\Gamma';\al,\al')  \nonumber \\
&\hspace{15mm} - I(k,k';\Gamma,\Gamma')\delta_{\al\al'} \big]^2, 
\end{align}
This is shown as error bars in Fig.~\ref{fig:svo} of the main text.
However, our numerical data are accurate enough and the errorbars are invisible.

\subsection{Time-reversal and inversion symmetries}
The time-reversal operation is performed by the antiunitary operator $\mathscr T$, which transforms the electron operator as
\begin{align}
\mathscr T c_{m\sg} \mathscr T^{-1} &= (-1)^{
m + \frac 1 2 -\sg } c_{-m,-\sg}
, \label{eq:TR_opration}
\end{align}
where $\sg = 1/2$ for $\ua$ spin and $\sg=-1/2$ for $\da$ spin.
The integer $m$ represents the magnetic quantum number which is an eigenvalue of $\hat L_z$ [in Eq.~\eqref{eq:TR_opration}, depending on the definition of spherical harmonics, the factor $(-1)^\ell$ is needed].
The complex number is also transformed as $\mathscr T z \mathscr T^{-1} = z^*$.
If the time-reversal symmetry is present, one can show the relation
\begin{align}
I(k,k';\Gamma,\Gamma')  &= (-1)^{k+k'} I(k,k';\Gamma,\Gamma')
,
\end{align}
which means that the odd-rank and even-rank multipoles do not couple.

The inversion $\mathscr I$ is performed for the operators as
\begin{align}
\mathscr I c_{m\sg} \mathscr I^{-1} = (-1)^\ell c_{m\sg}
.
\end{align}
Hence, for a fixed-$\ell$ subspace, there is no constraint on the interaction parameters.
On the other hand, if the parity mixing is considered, the inversion symmetry prohibits the coupling between even and odd angular momenta.

\begin{figure}[t]
\begin{center}
\includegraphics[width=60mm]{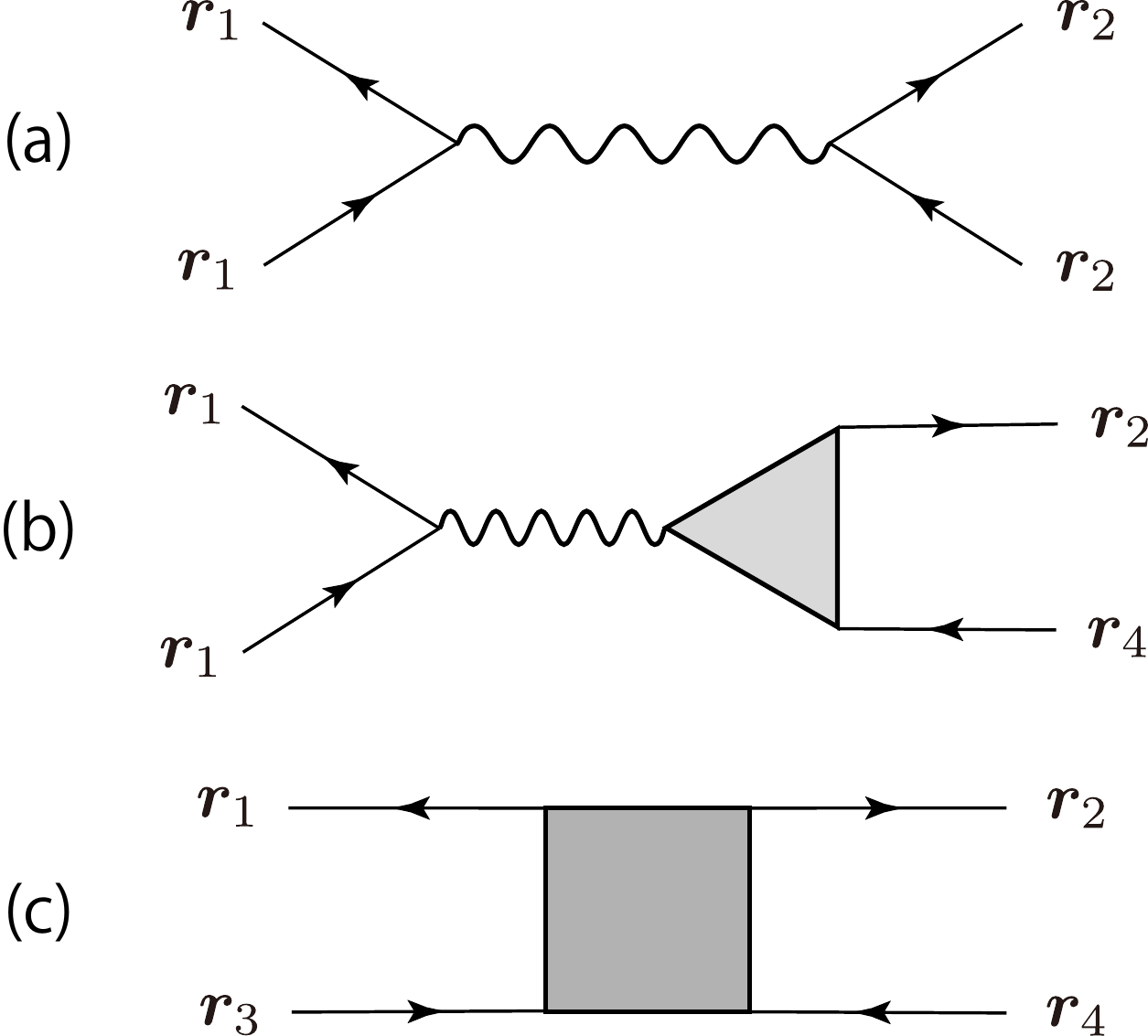}
\caption{
Schematic pictures for 
(a) $U(\bm r_1,\bm r_2,\bm r_1,\bm r_2)$,
(b) $U(\bm r_1,\bm r_2,\bm r_1,\bm r_4)$, and
(c) $U(\bm r_1,\bm r_2,\bm r_3,\bm r_4)$.
The solid lines with arrows show electron creation (outgoing) and annihilation (ingoing).
}
\label{fig:diagram}
\end{center}
\end{figure}

\subsection{Coulomb interaction with a specific form}
As discussed in the main text, the odd-rank multipoles are related the complexity of the spatial structure of the interactions.
To clarify this point, 
let us consider the following interaction form:
\begin{align}
\mathscr H_{\rm int} &= \frac 1 2 \int \diff \bm r_1  \diff \bm r_2 \diff \bm r_4
\nonumber \\
&\ \ \times U(\bm r_1,\bm r_2,\bm r_1, \bm r_4)
:n(\bm r_1,\bm r_1) n(\bm r_2,\bm r_4):
. \label{eq:int_spec}
\end{align}
The schematic illustration of this interaction is shown in Fig.~\ref{fig:diagram} together with the other types of interactions.
We expand the interaction function and field operators by the basis functions as
\begin{align}
\psi_{\sg}(\bm r) &= \sum_{n\ell m} R_{n\ell}(r) Y_\ell^m (\Omega) c_{n\ell m\sg}
, \\
U(\bm r_1,\bm r_2,\bm r_1,\bm r_4) &= \sum_{k m} \mathcal U_{k}(r_1; \bm r_2,\bm r_4) Y_{k}^m (\Omega_1)
,
\end{align}
where we have used the spherical coordinates $\bm r = (r, \Omega)$, $Y$ is the spherical harmonics, and $n$ is a principal quantum number.
We restrict ourselves to a single combination of $(n,\ell)$, which is to be omitted in the expression, 
and we obtain the interaction tensor of the form
\begin{align}
&U_{m_1m_2m_3m_4} = \sum_{k} \mathcal F_{k}^{m_1-m_3, m_2,m_4}
C_{k}(m_1,m_3)
\\
&C_k(m,m') = (-1)^m \int \diff \Omega \, Y_{\ell}^{-m}(\Omega) Y_{\ell}^{m'}(\Omega) Y_k^{m-m'}(\Omega)
,
\end{align}
where $\mathcal F$ is obtained by integrating $\mathcal U$ with basis functions.
The Gaunt coefficient $C_k$ has a finite value only when $k$ is an even number in the present case.
However, the even number of $k$ does not immediately mean the absence of the odd-rank multipole interactions in terms of $I(\xi,\xi')$, although they are related.

The important feature is that the material specific information is included only in the coefficient $\mathcal F$, which is a function of ``relative coordinate'' $m-m'$.
This fact motivates us to separate the sum into the combination of ``center-of-gravity coordinate'' ($M=\tfrac{m+m'}{2}$) and ``relative coordinate'' ($\mu = m-m'$) sums.
Then the multipole interactions are expressed as
\begin{align}
I(\xi,\xi')&= \frac{1}{(2\ell+1)^2}
\sum_{m_1m_2m_3m_4} U_{m_1m_2m_3m_4}O^\xi_{m_3m_1} O^{\xi'}_{m_4m_2}
\\
&= \frac{1}{2\ell+1}
\sum_{k\mu} \sum_{m_2m_4} \mathcal F_{k}^{\mu,m_2,m_4} G_{k\xi}^\mu O^{\xi'}_{m_4m_2}
,
\end{align}
where 
\begin{align}
G_{k\xi}^\mu &= \frac{ 1 }{2\ell+1} \sum_M 
C_k \big( \tfrac{2M+\mu}{2}, \tfrac{2M-\mu}{2} \big) O^\xi_{\frac{2M-\mu}{2}, \frac{2M+\mu}{2}}
.
\end{align}
Note that $G$ depends on the choice of matrix basis, but not on specific details of materials.
We have numerically confirmed for a given set of $\hat O^\xi$ for the point groups listed in Tabs.~\ref{tab:decomp} and \ref{tab:decomp2}
that $G_{k\xi}^\mu$ is finite only when the even number $k$ coincides  with the rank of $\xi$.
Since the multipole matrices for the other point groups are constructed from the linear combinations of the ones for the cubic group at each rank, $G_{k}$ for odd $k$ is generally zero.
When the time-reversal symmetry is preserved, the even- and odd-rank multipoles do not mix, and hence the interaction with the form \eqref{eq:int_spec} include only the even-rank multipoles in the interaction.

We contrast the above results with the more simplified interaction including $U(\bm r_1,\bm r_2,\bm r_1,\bm r_2)$ only.
The multipole interaction parameter is written as
\begin{align}
I(\xi,\xi')&= 
\sum_{k\mu}\sum_{k'\mu'} \mathcal F_{kk'}^{\mu\mu'} G_{k\xi}^\mu G_{k'\xi'}^{\mu'} 
.
\end{align}
With this expression, since $G$ is finite only if $\xi$ represents even-rank multipoles, the odd-rank multipole is absent regardless of the presence or absence of the time-reversal symmetry.
If we further use the condition
\begin{align}
\mathcal F_{kk'}^{\mu\mu'} &= \frac{4\pi (-1)^\mu F^k}{2k+1} \, \delta_{\mu,-\mu'}\delta_{kk'}
,
\end{align}
where $F^k$ is the Slater-Condon parameter, the results for the spherical limit 
with $U(\bm r,\bm r',\bm r,\bm r') \propto 1/|\bm r - \bm r'|$
 is recovered.
In this case the parameters $I$ and $F$ are connected by a simple relation.
We define the proportional constant by
\begin{align}
I(k,k) = H_k F^{k}
\end{align}
for a fixed $\ell$.
The values of $H$ are listed in Tab.~\ref{tab:corre}.

\begin{table}[t]
\centering
\begin{tabular}{cccccc}
\hline
$H_k(\ell)$ & \  &\ $k=0$\ &\ $k=2$\ &\ $k=4$\ &\ $k=6$\ \\
\hline
$\ell = 0$ ($s$-electron) && 1 & &  &   \\[1mm]
$\ell = 1$ ($p$-electron) && 1 & $\frac{2}{25}$ &  &   \\[1mm]
$\ell = 2$ ($d$-electron) && 1 & $\frac{2}{35}$ & $\frac{2}{63}$ &   \\[1mm]
$\ell = 3$ ($f$-electron) && 1 & $\frac{4}{75}$ & $\frac{2}{99}$ & $\frac{100}{5577}$  \\[1mm]
\hline
\end{tabular}
\caption{
Proportional constants $ H_k$ between multipole interaction $I(k,k)$ and Slater-Condon parameter $F^k$.
}
\label{tab:corre}
\end{table}

\section*{SM C. Details of the \textit{ab initio} calculation }

Details of the \textit{ab initio} calculation for SrVO$_3$ are as follows. 
The band structure calculation is based on the full-potential linear muffin-tin orbital (LMTO) implementation \cite{methfessel}. The exchange correlation functional is obtained by the local density approximation (LDA) of the Ceperley-Alder type and spin polarization is neglected.
We take the lattice constants of SrVO$_3$ as $3.844$ \AA.
The LDA calculation is done for the $12 \times 12 \times 12$ $\bm k$ mesh.
The angular momentum of the atomic orbitals is taken into account up to 4 for all the atoms.
In the cRPA calculation, $6 \times 6 \times 6$ $\bm k$ mesh is employed.
We construct 5 maximally localized Wannier functions having the V anti-bonding $3d$ orbitals hybridized with O-$2p$ orbitals from the 16 Kohn-Sham bands, where we exclude the O bonding bands.

For UBe$_{13}$, we have used VASP \cite{Kresse96} with the generalized gradient approximation (GGA) method.
The lattice constant is chosen as 10.268 \AA\ 
where two uranium atoms are included inside the unit cell.
The number of $\bm k$-point is $6\times 6\times 6$ in the Brillouin zone.
We construct the 136 Wannier orbitals (Be: $s,p$, U: $s,p,d,f$) from the 192 bands near the Fermi level.

\section*{SM D. Construction of $f^2$ wave functions}
In the main text, for $f^2$ wave functions, we treat the cubic deviation $\delta \mathscr H_{\rm C} = \mathscr H_{\rm C} - \mathscr H_{\rm C}^{\rm spher}$ from the spherical Coulomb interaction as a perturbation.
The unperturbed wave functions for $f^2$ configuration realized in Pr and U materials are constructed through the Hund's rules.
With a consideration of the spherical Coulomb interaction, the ground state is written by the total angular momentum $L=5$ and spin $S=1$ states as
\begin{align}
|L_zS_z\ra &= 
\sum_{mm'\sg\sg'} \la 3m3m'|LL_z\ra \la \tfrac 1 2 \sg \tfrac 1 2 \sg'|SS_z\ra
c^\dg_{m\sg} c^\dg_{m'\sg'} |0\ra 
\end{align}
for $f$ electrons ($\ell = 3$).
Here $\la J_{1}J_{1z} J_{2}J_{2z} | J_{3}J_{3z} \ra$ is the Clebsch-Gordan coefficient and we have introduced the vacuum $|0\ra$.
We consider the spin-orbital coupling and the resultant ground state is the 9-fold multiplet with the  total angular momentum $J=4$:
\begin{align}
|M\ra &= \sum_{L_zS_z} \la LL_zSS_z | JM\ra |L_zS_z\ra
,
\end{align}
which is to be normalized.
Then, using the first-order perturbation theory, the energy shift by the cubic deviation of the Coulomb interaction is obtained through the matrix element $\la M | \delta \mathscr H_{\rm C} |M'\ra$.
Under the cubic symmetry, it is convenient to move to the eigenfunction basis of the crystalline field Hamiltonian as follows:
\begin{subequations}
\begin{align}
|\Gamma_1
 \ra &= \sqrt{\tfrac{5}{24}} |4\ra + \sqrt{\tfrac{7}{12}} |0\ra + \sqrt{\tfrac{5}{24}} |-4\ra
\\
|\Gamma_3
,{1} \ra &= \sqrt{\tfrac{7}{24}} |4\ra - \sqrt{\tfrac{5}{12}} |0\ra + \sqrt{\tfrac{7}{24}} |-4\ra
\\
|\Gamma_3
,{2} \ra &= \sqrt{\tfrac{1}{2}} |2\ra + \sqrt{\tfrac{1}{2}} |-2\ra
\\
|\Gamma_4
,\pm \ra &= 
\sqrt{\tfrac{1}{8}} |\mp 3\ra + \sqrt{\tfrac{7}{8}} |\pm 1\ra
\\
|\Gamma_4
,0 \ra &= \sqrt{\tfrac{1}{2}} |4\ra - \sqrt{\tfrac{1}{2}} |-4\ra
\\
|\Gamma_5
,\pm \ra &= \sqrt{\tfrac{7}{8}} |\pm 3\ra - \sqrt{\tfrac{1}{8}} |\mp 1\ra
\\
|\Gamma_5
,0 \ra &= \sqrt{\tfrac{1}{2}} |2\ra - \sqrt{\tfrac{1}{2}} |-2\ra
\end{align}
\end{subequations}
Here $\Gamma_i$ is the Bethe symbol which is conventionally used for $f$-electron states.

\section*{SM E. Multipole representation of general interactions}
For a spin-orbital coupled basis, the interaction can in general be written as
\begin{align}
\mathscr H_U &= \frac 1 2 \sum_{ijkl} U_{ijkl} c^\dg_i c^\dg_j c_l c_k
.
\end{align}
Without loss of generality, the interaction tensor has the symmetry
\begin{align}
U_{ijkl} &= -U_{ijlk} = -U_{jikl} = U^*_{klij}
,
\end{align}
which originates from Hermiticity of the Hamiltonian and anticommutation relation of fermion operators.
Once we construct the complete set of Hermitian matrix basis \cite{Kusunose20,Wang17,Tamura20}, the multipole interactions are obtained as
\begin{align}
\mathscr H_U &= \frac 1 2 \sum_{\xi\xi'} I(\xi,\xi') : M^\xi M^{\xi'} :
, \\
M^\xi &= \sum_{ij}c_i^\dg O^\xi_{ij} c_j
, \\
I(\xi,\xi') &= \frac{1}{N^2}\sum_{ijkl} U_{ijkl} O^\xi_{ki} O^{\xi'}_{lj}
,
\end{align}
where $N=\sum_i 1$.

As an exemplary demonstration, let us consider the Hubbard interaction for spin-$1/2$ electrons:
\begin{align}
\mathscr H_U &= U c_\ua^\dg c_\da^\dg c_\da c_\ua
.
\end{align}
Its symmetrized form is
\begin{align}
&\mathscr H_U = \frac 1 2 \sum_{\sg_1\sg_2\sg_3\sg_4} U_{\sg_1\sg_2\sg_3\sg_4}
c^\dg_{\sg_1}c^\dg_{\sg_2} c_{\sg_4} c_{\sg_3}
, \\
&U_{\sg_1\sg_2\sg_3\sg_4} = \frac{U}{2} (\delta_{\sg_1\sg_3} \delta_{\sg_2\sg_4} - \delta_{\sg_1\sg_4} \delta_{\sg_2\sg_3})
.
\end{align}
The complete matrix basis in this case is the $2\times 2$ Pauli matrices and corresponding multipole operators are charge and spin:
\begin{align}
M^{\mu} &= \sum_{\sg\sg'} c^\dg_\sg \sg^\mu_{\sg\sg'} c_{\sg'}
\end{align}
for $\mu=0,1,2,3$.
The multipole representation is written as
\begin{align}
\mathscr H_U &= \frac U 8 \Big[
:(M^0)^2: - \sum_{\mu=1}^3 :(M^\mu)^2:
\Big] 
.
\end{align}
Thus we obtain symmetric expression with charge and spin operators.
Physically this expression indicates that, for $U>0$, the charge increases the energy while the emergence of the spin moment is energetically favorable.

\vspace{10mm}
\noindent
{\bf \large References}
\\[1mm]
See the list of references in the main text.


\begin{thebibliography}{99}
\bibitem{comment1} The operator expansion is in principle possible also for the spin-orbital coupled interactions. See SM E \cite{suppl}.
\end{thebibliography}
\end{document}